\documentclass{aastex}
\usepackage{emulateapj5}
\usepackage{epsf}
\usepackage{apjfonts}
\usepackage{amsmath,amsfonts}

\usepackage{graphicx}

\usepackage{natbib} 
\bibpunct{(}{)}{;}{a}{}{,} 

\newcommand{\myemail}{freitag@tapir.caltech.edu}

\newcommand{\SgrA}{Sgr~A$^\ast$}
\newcommand {\MBH}{\ensuremath{M_{\mathrm{BH}}}}
\newcommand {\RS}{\ensuremath{R_{\mathrm{S}}}}
\newcommand {\kms}{\ensuremath{\mathrm{km\,s}^{-1}}}
\newcommand{\gr}{gravitational radiation}
\newcommand{\gw}{gravitational waves}

\newcommand{\rem}[1]{} 

\shorttitle{Gravitational waves from Galactic center}
\shortauthors{Freitag}

\begin{document}


\title{Gravitational waves from stars orbiting the Sagittarius A$^\ast$ black hole}


\author{Marc Freitag\altaffilmark{1}}

\altaffiltext{1}{California Institute of Technology, Mail Code 130-33, Pasadena, CA 91125 USA, \myemail}




\begin{abstract}

One of the main astrophysical processes leading to strong emission of
gravitational waves to be detected by the future space-borne
interferometer {\em LISA} is the capture of a compact star by a black
hole with a mass of a few million solar masses in the center of a
galaxy. In previous studies, main sequence stars were thought not to
contribute because they suffer from early tidal disruption.  Here we
show that, according to our simulations of the stellar dynamics of the
{\SgrA} cluster, there must be one to a few low-mass main sequence
stars sufficiently bound to the central Galactic black hole to be
conspicuous sources in {\em LISA} observations. The probability that a
white dwarf may be detectable is lower than $0.5$ and, in spite of
mass segregation, detection of a captured neutron star or stellar
black hole in the center of the Milky Way is highly unlikely.

\end{abstract}




\section{Introduction}

Recently, the presence of massive black holes (MBHs, with masses from
a few $10^5$ to a few $10^9 M_\odot$) in the center of non-active
galaxies has received considerable observation support, mainly from
the kinematics of stars or gas in the center-most region of nearby
galaxies \citep{KG01}. One of the most convincing cases is the Milky
Way for which the motion of individual stars can be observed, both
along the line of sight or perpendicular to it
\citep{GPEGO00,GMBTK00,SchodelEtAl02}. They orbit a dark mass of
$2.6\times 10^6\,M_\odot$ whose extent is less
than 0.001\,pc and coincides with the radio source {\SgrA}.

These MBHs reside at the center of stellar clusters whose
densities may exceed $10^7\,M_\odot\,\mathrm{pc}^{-3}$
\citep{Lauer98,Alexander99}. An exciting consequence of this
is the possibility for a star to be captured onto a relativistic orbit
around the MBH by emission of {\gr}
\citep{HB95,SR97,Freitag01,Ivanov02}. For a source at a
distance of a few hundreds of Mpc and a MBH's mass in the range
between $10^4$ and $10^7\,M_\odot$, the frequency of these waves will
enter {\em LISA}'s range as the orbit shrinks down
\citep{Thorne98,Danzmann00}. If it
withstands the tidal forces, the star will eventually find itself on
an unstable, plunging orbit and disappear through the BH's
horizon. The computation of the late, strong-field, phases of this
inspiral requires full general relativistic treatment and cannot be
handled with tools presently available (see
\citealt{GK02,GHK02} and references therein). Nevertheless when the
pericenter distance, $R_\mathrm{p}$ is still well beyond the stability
limit ($3R_\mathrm{S}=6 G M_\mathrm{BH}c^{-2}$ for a circular orbit
around a non-rotating BH, with $G$ the gravitational constant,
$M_\mathrm{BH}$ the BH's mass and $c$ the speed of light), approximate
methods can be used to determine how the orbit evolves through
emission of {\gw}. For a star with mass $M_\ast$ on an initial orbit
of semi-major axis $a_0$ and high eccentricity, $1-e_0
\ll 1$, the time to plunge is
approximately \citep{Peters64}
\begin{eqnarray}
 T_\mathrm{plunge} &\simeq& \frac{2^{1/2}24}{85} \frac{c^5}{G^3
 M_\mathrm{BH}^2 M_\ast} (1-e_0)^{7/2} a_0^4 \simeq
 3.2\times 10^6\,\mathrm{yrs} \\
 && \nonumber \times
 \left(\frac{\MBH}{10^6\,M_{\odot}}\right)^{2}
\left(\frac{M_{\ast}}{1\,M_{\odot}}\right)^{-1}  
\left(\frac{R_\mathrm{p}^0}{10\,\RS}\right)^{4}  
\left(\frac{1-e_0}{10^{-5}}\right)^{-1/2},  
\end{eqnarray}
where $R_\mathrm{p}^0$ is the initial pericenter distance. Compact
stars, i.e. white dwarfs (WDs), neutron stars (NSs) or stellar black
holes (SBHs) spiral all the way down to the horizon but a main
sequence star (MSS) is torn apart by the tidal gravitational field of
the MBH at a distance $ R_{\mathrm{d}} =
k(\MBH/M_{\ast})^{1/3}R_{\ast} \simeq 50\,k\,\RS
(\MBH/10^6\,M_{\odot})^{-2/3}
(\bar{\rho}_\ast/1\,\mathrm{g}\,\mathrm{cm}^{-3})^{-1/3}$, where
$R_{\ast}$ is the radius of the star, $\bar{\rho}_\ast$ its average
density and $k$, a constant of order unity, depends on the stellar
structure \citep{Hills75,Rees88} . Hence a MSS may come close enough
to the MBH to emit significant amounts of gravitation radiation only
if it is compact enough. Note that $\bar{\rho}_\ast$ is maximum for
$M_\ast\simeq 0.07\,M_{\odot}$, at the transition to brown dwarfs
\citep{CB00}.

We only consider quadrupolar {\gr} because it dominates energy and
angular momentum losses. In the weak field approximation and
neglecting the gravitational influence of other stars, the orbit may
be treated as a Keplerian ellipse whose parameters slowly change as a
result of emission of {\gr}. The latter is emitted at integer
multiples of the orbital frequency, $\omega_n=n (G\MBH)^{1/2}a^{-3/2}$.
At a distance $D$ from the source, the strain amplitude in the
$n^\mathrm{th}$ harmonic is
\begin{eqnarray}
	h_n &=& \gamma(n,e) \frac{1}{D} \frac{G^2\MBH M_{\ast}}{c^4a} \\
        \nonumber &\simeq& 2\times 10^{-27} \gamma 
	\left(\frac{D}{1\,\mathrm{Mpc}}\right)^{-1}
	\left(\frac{a}{1\,\mathrm{pc}}\right)^{-1}
	\left(\frac{\MBH}{10^6\,M_{\odot}}\right)
	\left(\frac{M_\ast}{1\,M_{\odot}}\right),
\end{eqnarray}
where the non-dimensional factor $\gamma$ is an intricate function of
$n$ and the eccentricity $e$ \citep{PM63,PPSLR01}. For simplicity, we
consider the the {\em rms amplitude}, averaged over $+$ and
$\times$ polarizations and all directions. 

We assume that a star is irremediably captured by the MBH if it gets
on an orbit with a time scale for shrinkage by emission of {\gr}
shorter than the time over which 2-body relaxation could significantly
modify the pericenter distance,
$T_{\mathrm{mod}}\simeq \theta^2 T_\mathrm{rel} < T_\mathrm{rel}$ \citep{SR97,Ivanov02}.
$T_\mathrm{rel}$ is the usual relaxation time \citep{BT87} and
$\theta$ is the angle between the trajectory and the direction to the
center. The typical eccentricity of capture orbits may be estimated
from the condition $T_{\mathrm{plunge}} < T_\mathrm{rel}$:
\begin{equation}
	1-e < 
	\left(\frac{V_{\mathrm{orb}}}{c}\right)^{\!\!10/7} \!\!
	\left(\frac{\MBH}{M_{\mathrm{cusp}}}\right)^{\!\!2/7} \simeq 
         10^{-5} \left(\frac{V_{\mathrm{orb}}}{100\,\kms}\right)^{\!\!10/7}.
\end{equation}
Here, $V_{\mathrm{orb}}$ is a typical velocity for stars in the
``cusp'', i.e. the central region of the cluster where the kinematics
is dominated by the gravitational attraction of the MBH.
$M_{\mathrm{cusp}}$, of order \MBH, is the total stellar mass in this
region. The rate of capture is controlled by processes that send
stars onto these extremely elongated orbits. In a spherical cluster,
the most important one is 2-body relaxation.

\section{Numerical model of the central cluster}

To compute capture rates, many difficulties have to be faced, in
particular the role of mass segregation, stellar collisions or tidal
disruptions. Although these complications can be included to some
extent into analytical computations \citep{SR97,MEG00}, the intricate
nature of the problem calls for numerical simulations. We have recently
developed a new Monte Carlo (MC) code to follow the long term
evolution of galactic nuclei \citep{FB01a,FB02b,Freitag01}. This tool is
based on the scheme proposed by \citet{Henon73} to simulate
globular clusters but, in addition to relaxation, it also includes
collisions, tidal disruptions, stellar evolution and captures.  \rem{The MC
technique assumes that the cluster is spherically symmetric and in
dynamical equilibrium. It represents it as a set of particles, each of
which may be considered as a homogeneous spherical shell of stars
sharing the same orbital and stellar properties.}


For this work, we devised a model which represents the central cluster
of our Galaxy. It is an $\eta$-model \citep{Tremaine94}. The stellar
density is
$\rho_\ast(R) = \eta M_{\mathrm{cl}}/(4\pi R_\mathrm{b}^3)\,
  r^{\eta-3}(1+r)^{\eta-1}$,
with $r=R/R_\mathrm{b}$, $\eta=1.3$, $M_\mathrm{cl} = 8.67\times 10^7\,M_\odot$ (total
stellar mass) and $R_\mathrm{b} = 22$\,pc. The central BH has a mass
of $\MBH=2.6\times 10^6\,M_\odot$. These quantities were chosen to
provide a good agreement with the observed run of the enclosed mass
around {\SgrA} as a function of the distance and the star
counts \citep{GPEGO00}, while imposing
$\mu\stackrel{\mathrm{def}}{=}\MBH/M_{\mathrm{cl}}=0.03$, a relatively
large ratio chosen for the sake of resolution. We assume all stars
formed $10^{10}$\,years ago with a ``universal'' initial mass function
(IMF), ${\mathrm{d}N_{\ast}}/{\mathrm{d}M_{\ast}}
\propto M_{\ast}^{-\alpha}$ with $\alpha=0.3$ between $0.01$ and
$0.08\,M_\odot$, 1.3 between $0.08$ and $0.5\,M_\odot$ and 2.3 up to
$120\,M_\odot$ \citep{Kroupa00b}. We do not include giant stars. All WDs,
NSs and SBHs are assumed to have 0.6, 1.4 and 7\,$M_\odot$,
respectively. There is no initial mass segregation. The number of
particles for our main simulation is $6\times 10^6$, so that each
particle represents 65.5 stars.

\section{Results of simulations}

\begin{figure*}

   \begin{center}
  \resizebox{0.5\hsize}{!}{\includegraphics{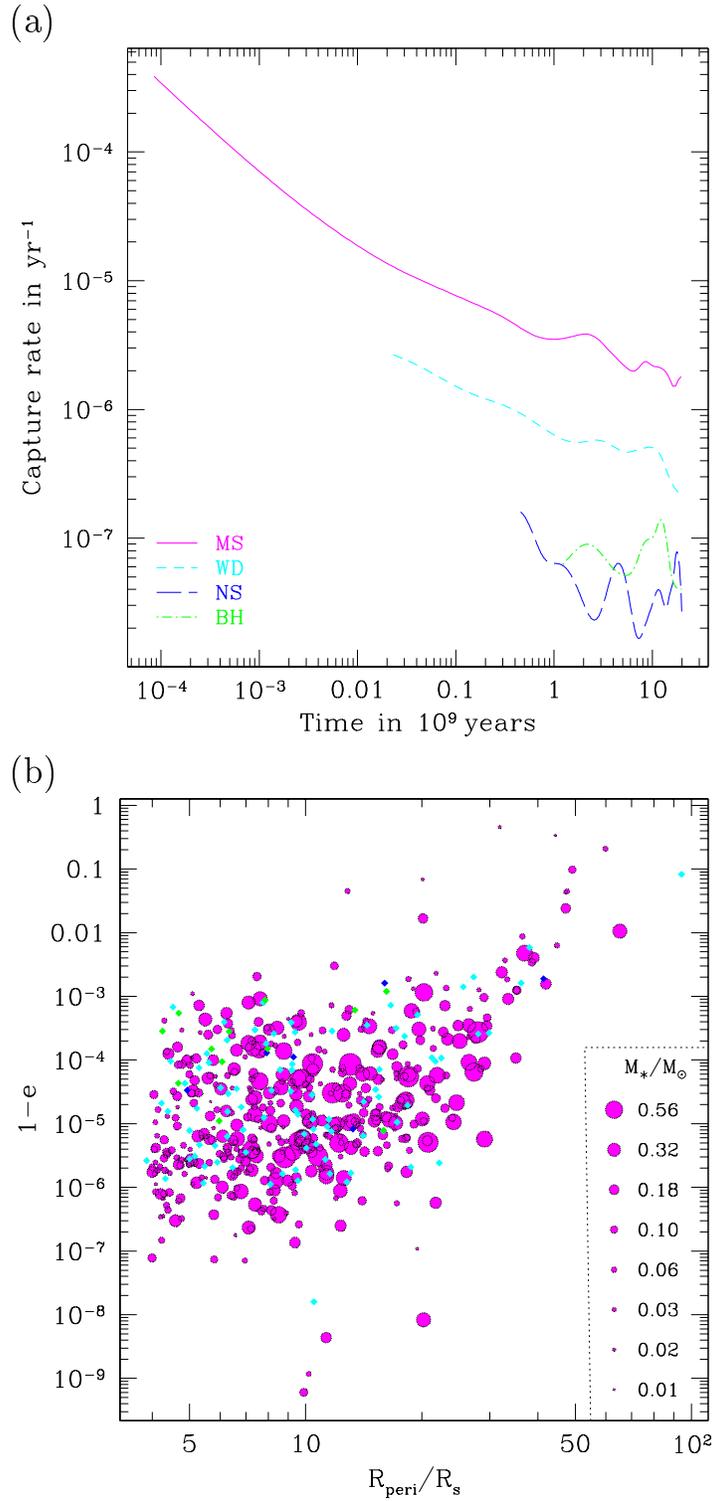}}
    \end{center}

  \caption{ Captures through emission of gravitational waves
  for our Monte Carlo simulation of the Galactic center. {\em
  (a)} Evolution of the capture rates for the various stellar
  species. Note that only a small number of events have
  occurred for SBHs or NSs, hence the noisy curves. {\em
  (b)} Orbital parameters at capture for each event
  (which has a statistical weight of 65.5 stars). $e$ is the
  eccentricity and $R_\mathrm{p}$ the pericenter distance (in units
  of the Schwarzschild radius, $R_\mathrm{S} =
  2.5\times10^{-7}$\,pc). Only events that occurred in
  the first 10\,Gyrs of the simulation are included. The surface of
  disks for MSSs is proportional to the mass of the star. Capture of
  compact remnants are represented with diamonds.}

\label{fig:capt}
\end{figure*}

In Fig.~\ref{fig:capt}a, we show the capture rates as a function of
time. As captures are rare, we have integrated the evolution of the
nucleus over many billion years to improve statistics. However, over
such a long time scale, the nucleus model experiences a relatively
important evolution. Most notably, significant mass segregation
occurs. At the end of the simulation, the density is dominated by SBHs
(which represent only 2\,\% of the stellar mass) in the region
interior to $\sim 0.15\,$pc while the density of MSSs has decreased
from a steep $\rho_\ast
\propto R^{-1.7}$ to a milder $R^{-1.3}$ peak. This explains the steady decrease in the
MSS capture rate. Given its efficiency, mass segregation should in
principle be introduced from the beginning of the simulation, which is
unfortunately impossible for lack of observational constrains and
adequate cluster models. Anyway, it is clear that the predicted MSS
capture rate is in excess of $2\times 10^{-6}\,\mathrm{yr}^{-1}$ and
may be as high as a few $10^{-5}\,\mathrm{yr}^{-1}$. For WDs, the rate
is a few $10^{-7}\,\mathrm{yr}^{-1}$ and around $5\times
10^{-8}\,\mathrm{yr}^{-1}$ for NSs and SBHs. These latter values are
based on a small number of events and have low statistical
significance. In Fig.~\ref{fig:capt}(b), the initial orbital
parameters are reported for all captures. One sees that orbits are
very elongated, as predicted, with very small pericentre distances.
Previous, analytical, studies did not address the capture of
MSSs. They predicted a WD capture rate ranging from
$10^{-8}\,\mathrm{yr}^{-1}$ \citep{HB95,Ivanov02} to
$10^{-7}\,\mathrm{yr}^{-1}$ \citep{SR97}. Explaining why these
estimates are different from each other and from our result is beyond
the scope of this letter as it would require a lengthy discussion of
the different nucleus models and various treatments of the physics in
these papers. We note that our WD rate is similar but larger than that
of \citet{SR97} while these authors predict an initial SBH capture rate
some three orders of magnitude larger than ours!

\begin{figure*}

   \begin{center}
  \resizebox{0.5\hsize}{!}{\includegraphics[bb=20 151 562 689,clip]{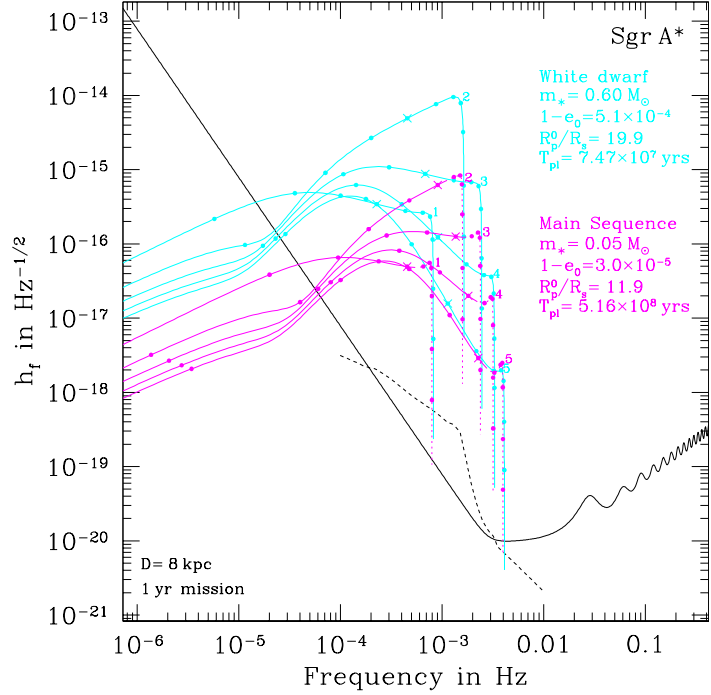}}
    \end{center}

  \caption{ Gravitational signal for two events from our \SgrA
  simulation, a WD and a low-mass MSS. We plot the amplitude versus
  frequency for the 5 first Fourier components of the quadrupolar
  radiation \citep{PPSLR01}. We use $h_f \stackrel{\mathrm{def}}{=}
  h^\mathrm{rms}\sqrt{T_\mathrm{obs}}$ where $T_\mathrm{obs}$ is the
  effective duration of observation at frequency $f$, i.e., the
  mission duration (one year) if the signal is stable over that
  period.  The sharp drop in the signal amplitude at the end of the
  evolution is due to either the time left before swallowing or the
  characteristic time for frequency increase,
  $f(\mathrm{d}f/\mathrm{d}t)^{-1}$, being shorter than the mission
  duration.  The crosses represent the position 1000 years before
  plunge through the horizon. Other ticks are for positions $10^6$,
  $10^5$, $10^4$, 100, 10, 1 year, 1 month and 1 day before
  plunge. The segments in dotted line for the MSS correspond to
  $R_{\mathrm{p}}<R_{\mathrm{d}}$, a regime where tidal
  disruption should have occurred.  The solid V-shaped curve
  represents {\em LISA}'s intrinsic noise \citep[$\mathrm{SNR}=1$,][]{LHH00}. The dashed
  line is an estimate of the confusion noise due to unresolved WD
  binaries in our Galaxy \citep{BH97}. }

\label{fig:ampl}
\end{figure*}

During the course of the MC simulations, captured stars are
immediately removed from the cluster. Their initial orbital parameters
being known, we come back to this data and compute the
inspiralling and concomitant {\gw} emission
\citep{PPSLR01,GHK02}. To illustrate this, in Fig.~\ref{fig:ampl}, we
plot the evolution of the frequency and amplitude of the lowest five
harmonics of the emitted waves for typical WD and MSS events. The WD would be
detectable during a few $10^5$\,years and the MSS for nearly
$10^6$\,years. NSs and a SBHs typically have shorter detectability
durations. Multiplying these times by the respective capture rates,
one gets rough estimates for the expected average number of {\em LISA}
sources at the Galactic center.  However, the MC data permits more
precise determinations through the following method. For each event
$i$, we computed the evolution of the 20 first harmonics of the rms
amplitude and selected at each time the one with the highest {\em
LISA} signal-to-noise ratio (SNR). Then we compute $T_i^{(\mathrm{detec})}(s)$, the time
during which there is a harmonic with SNR larger than $s$. Finally, we
get the expected number of sources with a harmonic component stronger
than SNR $s$ by summation,
\begin{equation}
  N^{(\mathrm{detec})}(s) = \frac{N_\ast}{N_\mathrm{p}} \frac{1}{\Delta T_{\mathrm{simul}}}
  \sum_{i=1}^{N_\mathrm{events}}\,T_i^{(\mathrm{detec})}(s).
\end{equation}
The summation is realized over all the $N_\mathrm{events}$ events that
occurred during some given time interval $\Delta
T_{\mathrm{simul}}$. 

Fig.~\ref{fig:sourcesnumber} is the result of this computation. To
have reasonable statistics, we set $\Delta T_{\mathrm{simul}}=
10^9$\,yrs for MSSs and WDs and $2\times 10^{10}$\,yrs for NSs and
SBHs. The most striking results concern MSSs. Even though complete
tidal disruption has been assumed when the particle reaches
$R_\mathrm{d}$, the predicted number of sources with $\mathrm{SNR}\ge
10$ is of order 3-5. If sources with SNR of 3 can be detected, then
{\em LISA} should be able to observe of order 10 MSSs orbiting the
{\SgrA} MBH. Note that for such long-lived sources, the SNR is
proportional to the square root of the mission duration, likely to
exceed one year.

\section{Discussion}

\begin{figure*}

   \begin{center}
  \resizebox{0.5\hsize}{!}{\includegraphics[bb=26 154 562 671,clip]{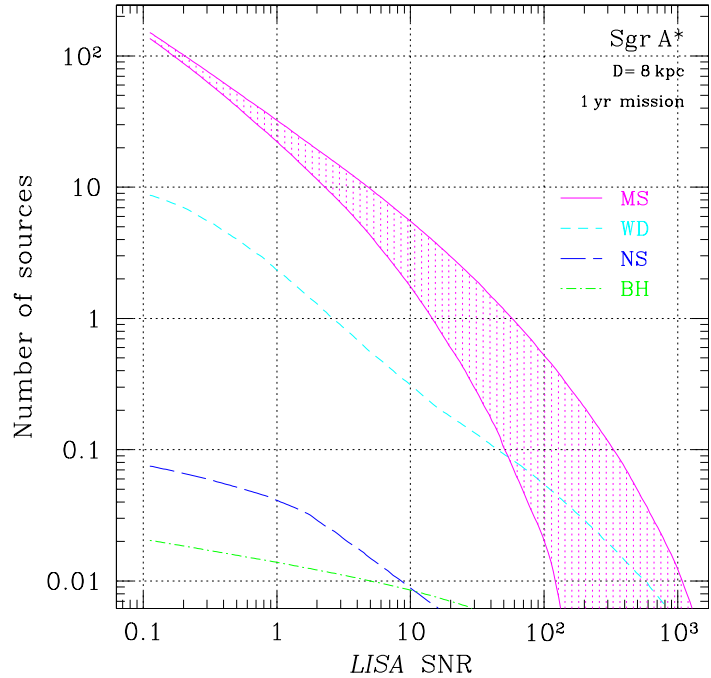}}
    \end{center}

  \caption{ Expected number of sources of {\gw} at the
  Galactic center. This figure shows the number of MSSs, WDs, NSs and
  SBHs predicted to produce a signal above a given signal-to-noise
  ratio ($\mathrm{SNR}$). For MSSs and WDs, the capture events that
  occurred in the first $10^9$\,years of our Monte Carlo simulation
  have been included in the computation (95 and 11 events
  respectively). For NSs and BHs, we used the first $2\times 10^{10}$\,years (13 and
  22 events). The orbital evolution of each captured star, as driven
  by emission of gravitational radiation around a non-spinning black
  hole, has been integrated down to plunge instability or tidal
  disruption \citep{GHK02} and, at each time, we select the
  Fourier component of the quadrupolar radiation \citep{PPSLR01}
  yielding the highest SNR. The upper curve for MSSs is obtained when
  tidal heating is neglected. The lower curve corresponds to a
  pessimistic estimate of the decrease in the number of sources due to
  tidal heating, as described in the main text (the number of
  contributing events is reduced to 27).}

  \label{fig:sourcesnumber}
\end{figure*}

A major concern regarding this result is the role of tidal
interactions which may enter the game before they are strong enough to
disrupt a star in a single passage. Tides are raised at each
pericentre passage, converting some orbital energy into stellar
pulsation. These oscillations dampen into heat through mechanisms and
on time scales that are still uncertain \citep{McMMcDT87,KG96}.
Conservatively, we assume complete dissipation, thus disregarding the
possibility of energy being transfered back from the oscillations to
the orbital motion at next periastron passage, and ineffective
radiation of this energy. The ratio of the amount of energy to be
dissipated (either by {\gw} or tides) to reach a relativistic
detectable orbit, to the star's self-binding energy is of order
$(\RS/R_\mathrm{p})(c/V_\ast)^2(1-e)
\gg 1$ where $V_\ast\approx 600\,\kms$ is the escape velocity from the stellar surface. 
Consequently, with our assumptions, tidal dissipation cannot influence
the orbit without first strongly affecting the stellar
structure. Hence, to get a (over)estimate of the impact of tidal
interactions in reducing the expected number of detectable stars, we
computed for all capture events the tidal energy pumped into the star
at each periastron passage, assuming $e\simeq 1$ and using the $n=1.5$
polytrope model from \citet{McMMcDT87}. When the accumulted energy
reached 20\,\% of the self-binding energy, the star was considered
disrupted. Recomputing $N^{(\mathrm{detec})}(s)$ with this pessimistic
procedure, we still get of order 0.5--2 and 4--8 MSS sources with
$\mathrm{SNR}\ge 10$ and $\mathrm{SNR}\ge 3$, respectively, see
Fig.~\ref{fig:sourcesnumber}. We conclude that tidal interactions can
potentialy reduce the number of sources significantly but probably not
supress them completely.

We have tested the robustness of our result to several other aspects
of the modeling, through a series of lower resolution simulations
($N_\mathrm{p}=10^6$). Interestingly, our conclusions are not affected
if we assume a constant stellar formation rate during the
$10^{10}\,$years preceding the beginning of the
simulation. Observations of the Arches cluster, close to the Galactic
center, are hinting to an IMF exponent of $\alpha\simeq 1.65$ for
$M_\ast>10\,M_\odot$ \citep{FKMSRM99}. If we adopt $\alpha=1.65$ for
stars more massive stars than $8\,M_\odot$, the capture rate is
slightly lower. With the extreme assumption that $\alpha=1.65$ above
$0.5\,M_\odot$, a very SBH-rich population is formed. After a few
billion years, they dominate the central pc from which they have
expelled most of the MSSs whose capture rate is reduced by a factor of
3. Modifications of the low-mass part of the IMF have direct
consequences because only stars less massive than $\sim 0.6\,M_\odot$
may be detected before tidal disruption. The IMFs with the largest or
lowest proportion of low mass stars that are still compatible with
local observations
\citep{Kroupa00b} lead to a number of predicted MSS sources higher by
40-80\,\%, respectively lower by of order 25\,\%.

Only low mass MSSs contribute as more massive but less denser ones
suffer from early tidal disruption. Hence, the MSS signal would be too
weak to be detected from distant galaxies. A MS source would only be
marginally detectable during of order $10^3$ years if located in the
Virgo cluster which is too short to ensure a significant detection
probability. Local group galaxies are probably ruled out as well. The
BH at the center of M~31 has a mass of a few $10^7\,M_\odot$ so that
the frequency of gravitational waves would be too low. Detailed
simulations for M~32 yield a probability for a MS source
with SNR above 1 of order 0.3--0.5 only. If massive BHs are present in
dwarf galaxies or globular clusters
\citep{GRH02,GerssenEtAl02} and their masses follow the well
known correlation with the velocity dispersion of the host galaxy
\citep{TremaineEtAl02}, they impose too intense tidal forces in their
vicinity. To summarize, captured MSSs should be detected at the
Galactic center, many $10^5$ years before plunge, but all other
sources are predicted to be compact remnants in galaxies at distances
of a few hundreds of Mpc, during the last few months or years of
inspiral.

\acknowledgments

The draft of this paper has been written during a visit to Princeton
University Observatory thanks to the hospitality of Laurent Eyer. This
work was realized under a grant from the Swiss National Science
Foundation. Complementary support from NASA under grant NAG5-10707 is
acknowledged. The author expresses his gratitude to Kostas
Glampedakis who provided the routines to compute the orbital
evolution, to Sterl Phinney for interesting discussions and to Scott
A. Hughes and Tal Alexander for useful comments on the draft of this
paper.



\end{document}